\documentstyle[prc,aps,epsfig]{revtex}
\begin{document}
\draft

\title{\bf New features of two-particle correlations}
\author{G.A.Kozlov}
\address{Bogoliubov Laboratory of Theoretical Physics,\\
Joint Institute for Nuclear Research, 141980 Dubna,
Moscow Region, Russia}

\date{\today}
\maketitle

\begin{abstract}
We show that a recently proposed derivation of Bose-Einstein
correlations (BEC) by means of thermal
Quantum Field Theory, supplemented by operator-field evolution
of Langevin type, allows a deeper understanding of a
possible coherent behaviour of the emitting source and a clear
identification of the origin of the observed shape of the 2-particle BEC
function $C_2(Q)$. We explain the origin of the dependence of the measured
correlation radius on the hadron mass. The lower bound of the particle
source size is estimated.
\end{abstract}

\vspace{1 cm}

In our previous papers [1,2] we have established the master equation
of the field-operator evolution (Langevin-like [3]) equation which allows
one to gain a better understanding of  possible coherent behaviour
of the emitting source of elementary particles.
A clear identification of the shapes of both two-particle both  Bose-Einstein
and Fermi-Dirac correlation functions was observed in LEP experiments
ALEPH [4], DELPHI [5] and OPAL [6] which also suggested a dependence of
the measured correlation radius on the hadron mass $m$.

 We focused [2,1] our attention on two specific
features of Bose-Einstein correlations (BEC) clearly visible when
BEC are presented in the language of Quantum Field Theory (QFT)
supplemented by the operator-field evolution.  The features we discussed are:
$(i)$ how does a possible
coherence of the deconfined or hadronizing systems (modelled here
by some external
stationary force occurring in the Langevin-like equations) describing
the process of deconfinement or hadronization influence
the 2-particle BEC function
$C_2(Q=\sqrt {(p_{\mu} - p^{\prime}_{\mu})^{2}})$, where two particles
are characterized by their four-momenta  $p_{\mu}$ and $p^{\prime}_{\mu}$, and
$(ii)$ what is the true origin of the experimentally
observed $Q$-dependence of the $C_2(Q)$ in the approach used in [2,1]?

The only physical meaning of the memory term $\tilde K(p_{\mu})$, the noise
spectral function $\psi (p_{\mu})$, and the so-called
coherence function $\alpha = \alpha (m^2,\vec p^2,\beta)$ related to them [1]
\begin{equation}
\alpha (m^2,\vec p^2,\beta)= \frac{M^{4}_{ch}\,\rho^{2}(\omega,\epsilon)\,P^2}
{{\vert\omega - \tilde{K}(p_{\mu})\vert}^{2}\,n(\omega,\beta)}
\,\,\, {\rm as}\,\, \epsilon > 0 \,\,\,\,{\rm and}\,\, Q^2 \rightarrow 0,
\label{e1}
\end{equation}
were not yet investigated carefully (here $\omega$ is the Fourier transformed component
of 4-vector $p_{\mu}$ and it is conjugated to time $t$).
We have already established [1] that at $p_{\mu} - p^{\prime}_{\mu}\neq 0$ there is a
finite volume $\Omega(r)$ of a two-particle emitter source, and the correlation picture does exist.
The space in which the massive fields are settled down has an own characteristic length
$L_{ch}\sim M^{-1}_{ch}$ in the same sense like the massive field with the mass $m_{c}$ for
the Compton length $\lambda_{c} = m^{-1}_{c}$ is served. The appearance of the $\delta$-like
distribution
$\rho(\omega,\epsilon)= \int dx\,e^{(i\omega -\epsilon)x}\rightarrow \delta (x)$ as
$\epsilon\rightarrow 0$ is the consequence of the infinite volume of produced particles
(formally, $\epsilon$ is given by the physical spectrum).
The volume $\Omega(r)$ is conjugate to
some characteristic mass scale $M_{ch}$, $\Omega(r)\sim 1/{M^{4}_{ch}}$ in such a way that
$M_{ch}\rightarrow\infty$ when the correlation domain
does not exist. Hence, $M_{ch}\rightarrow 0$ obeys the condition  $\alpha\rightarrow 0$
that means no distortion (under the constant force $P$) acting on the system of produced particles.
On the other hand,
any distortion which disturbs this system by any reason (the coherence function $\alpha > 0$)
leads to a finite
domain of produced particles ($M_{ch}\neq 0$). Rather a strong
distortion, $\alpha\rightarrow\infty$, satisfies the point-like region of the particle
source emitter.

It is a difficult task to derive both
$\tilde K(p_{\mu})$ and $\psi (p_{\mu})$ using the general properties
of the quantum picture. The fact which is worthwhile to stress is that
the knowledge of one of these functions leads to understanding of the other
one, because of the normalization condition [2]
\begin{equation}
\int^{+\infty}_{-\infty}\frac{d\omega}{2\,\pi}{\left\vert
\frac{\psi (p_{\mu})}{\omega - \tilde K(p_{\mu})}\right\vert}^{2} = 1,
\label{e2}
\end{equation}

which is nothing else but the generalized fluctuation-dissipation theorem.
The spectral properties
\begin{equation}
{\vert\psi (p_{\mu})\vert}^2\rightarrow\frac{2}{\omega}\,\sin(\beta\omega)
{\vert\omega - \tilde K(p_{\mu})\vert}^2,\,\,\, 0<\beta <\infty
\label{e3}
\end{equation}
 follow from (\ref{e2}) with
\begin{equation}
\lim_{\beta\rightarrow\infty} \frac{1}{\pi}\,\frac{\sin(\beta\omega)}{\omega}
= \delta(\omega).
\label{e4}
\end{equation}

In this paper we would like to focus our attention on the role of the hadron mass
 influencing correlations between particles. To solve this problem,
one has to derive the memory term $\tilde K(p_{\mu})$ using the general properties
of QFT.

It is supposed that we work with the fields corresponding to the thermal
field $\Phi(x)$ with the standard definition of the Fourier transformed propagator
$F[\tilde G(p)]$
\begin{equation}
F[\tilde G(p)]= G(x-y) = Tr\left\{T[\Phi(x)\Phi(y)]\rho_{\beta}\right\},
\label{e5}
\end{equation}
with $\rho_{\beta}= e^{-\beta H}/Tr e^{-\beta H}$ being the density matrix of a local
system in the equilibrium with the temperature $T=\beta^{-1}$ under the Hamiltonian $H$.

We consider the interaction of $\Phi(x)$ with the external scalar field which is
given by the potential $U$. This potential contrary to an electromagnetic
field is a scalar one, but not a component of the four-vector. The Lagrangian
density looks like
\begin{equation}
L(x) = \partial_{\mu}\Phi^{\star}(x)\partial^{\mu}\Phi(x) -
(m^2 + U)\Phi^{\star}(x)\Phi(x)
\label{e6}
\end{equation}
and the equation of motion is
\begin{equation}
(\nabla^2 + m^2)\Phi(x) = -J(x),
\label{e7}
\end{equation}
where $J(x) = U\Phi(x)$ is the source density operator. Such a simple model allows one
to investigate the origin of the occurrence of the condensate in the static restricted
potential of the external field. We are interested in the origin of the unstable state
of the thermalized equilibrium in the nonhomogeneous external field under the influence
of the source density operator $J(x) = U\Phi(x)$. For example, the source can be considered
as the $\delta$-like generalized function $J(x)=\tilde\mu\,\rho(x,\epsilon)\Phi(x)$, where
$\rho(x,\epsilon)$ is the $\delta$-like succession giving the $\delta$-function as
$\epsilon\rightarrow 0$ ($\tilde\mu$ is some massive parameter). This model is useful
because the $\delta$-like potential $U(x)$ provides the model conditions
to restrict the particle emission domain or the deconfinement region. We suggest the
following form
\begin{equation}
J(x) = - \Sigma(i\partial_{\mu})\,\Phi(x) + J_{R}(x),
\label{e8}
\end{equation}
where the source $J(x)$ is decomposed into the regular systematic motion part
$\Sigma(i\partial_{\mu})\,\Phi(x)$ and the random source $J_{R}(x)$. Thus, the
equation of motion (\ref{e7}) becomes
\begin{equation}
(\nabla^2 + m^2 - \Sigma(i\partial_{\mu}))\Phi(x) = -J_{R}(x),
\label{e9}
\end{equation}
and the propagator satisfies the following equation:
\begin{equation}
[-p^{2}_{\mu} +m^2 -\tilde\Sigma(p_{\mu})]\tilde G(p_{\mu}) =1.
\label{e10}
\end{equation}

Now let us introduce the general non-Fock representation of the canonical commutation
relation (CCR). To this, we consider the operator generalized functions
\begin{equation}
b(x) = a(x) + R(x),
\label{e11}
\end{equation}
\begin{equation}
b^{+}(x) = a^{+}(x) + R^{+}(x),
\label{e12}
\end{equation}
where the operators $a(x)$ and $a^{+}(x)$ obey the CCR-relations:
\begin{equation}
[a(x),a(x^{\prime})]= [a^{+}(x),a^{+}(x^{\prime})]= 0,
\label{e13}
\end{equation}
\begin{equation}
[a(x),a^{+}(x^{\prime})]= \delta (x-x^{\prime}).
\label{e14}
\end{equation}
The operator generalized functions $R$ carry random features
describing an action of the external forces.

Both $b^{+}$ and $b$ obviously define the CCR representation. For each function
$f$ from the space $S(\Re_{\infty})$ of smooth decreasing functions one can
establish new operators $b(f)$ and $b^{+}(f)$
\begin{equation}
b(f) = \int f(x) b(x) dx = a(f) +\int f(x) R(x) dx,
\label{e15}
\end{equation}
\begin{equation}
b^{+}(f) = \int \bar f(x) b^{+}(x) dx = a^{+}(f) +\int\bar f(x) R^{+}(x) dx.
\label{e16}
\end{equation}
The transition from the operators $a(x)$ and $a^{+}(x)$ to $b(x)$ and $b^{+}(x)$,
obeying those commutation relations as $a(x)$ and $a^{+}(x)$ leads
to linear canonical representations. If both $b(f)$ and $b^{+}(f)$ create the
Fock representation of the CCR, then one can find the operator $U$ obeying the
following conditions:
\begin{equation}
U a(f) U^{-1} = b(f),
\label{e17}
\end{equation}
\begin{equation}
U a^{+}(f) U^{-1} = b^{+}(f).
\label{e18}
\end{equation}

Now we are going to a simple physical pattern.
Let us define the differential evolution (in time) equation, where the
sharp and chaotically fluctuating function, obeying this equation,
is the main object. Since we deal with the continuous time, this allows
one to formulate the stochastic differential equation applied to each
(analytical) function under the distortion of a random force. In classical
mechanics, the stochastic processes in a dynamic system are under a
weak action of the "large" system [7]. The "small" and the "large" systems
are understood in the sense that the number of the states of freedom of
the first system is less than that of the other one. We do not exclude even
the interplay between these systems. In the case when the "large" system
is in equilibrium state (in may be a thermostat state) our method will
allow one to describe the approximation to the distribution of the
probability to find physical states in the "small" system. On the quantum
level the role of such a "small" system is played by the restricted region
of produced particles, a particle source emitter with a definite size,
which we study in this paper.

Following the idea of classical
Brownian motion [8,3] of a particle with a unit mass, a charge $g$ and velocity $v(t)$
in the external, let say, electric field $E$, one can write the formal equation
describing the evolution (in time) of this particle in the following form:
\begin{equation}
\partial_{t}v(t) = \gamma\,v(t) + F + g\,E,
\label{e19}
\end{equation}
where $F$ stands for a random force subject to the Gaussian white noise;
$\gamma$ is a friction coefficient.



Referring to [2,1] for details, let us recapitulate here the
main points of our approach in the quantum case. The collision process produces a lot of
particles out of which we select only one (we assume for simplicity that
we deal only with identical bosons) and describe it by stochastic
operators $b(\vec{p},t)$ and $b^{+}(\vec{p},t)$, carrying the features of
annihilation and creation operators, respectively (the notation is the usual
one: $\vec{p}$ is $3$-momentum and $t$ is a real time).
The rest of the particles
are then assumed to form a kind of heat bath, which remains in
equilibrium characterized by a temperature $T$ (which will be
one of our parameters).
We shall also allow for some external (to the above heat bath)
influence to our system.
The time evolution of such a system is then assumed to be given by a
Langevin-type equation [2,1] for the new stochastic operator $b(\vec{p},t)$
\begin{equation}
i\partial_t b(\vec{p},t) =  A(\vec{p},t) + F(\vec{p},t) + P
\label{e20}
\end{equation}
(and a similar conjugate equation for $b^{+}(\vec{p},t)$). We assume that
an asymptotic free undistorted operator is $a(\vec{p},t)$, and the deviation
from the asymptotic free state is provided by the random operator
$R(\vec{p},t)$: $a(\vec{p},t)\rightarrow b(\vec{p},t) = a(\vec{p},t) +
R(\vec{p},t)$. It means that, e.g., the number of particle density
(a physical number) ${\langle n(\vec{p},t)\rangle}_{ph} \langle n(\vec{p})\rangle + O (\epsilon)$, where ${\langle n(\vec{p},t)\rangle}_{ph}$
means an expectation value by a physical state, while ${\langle n(\vec{p})\rangle}$
denotes that by an asymptotic state. In case we ignore the deviation
from the asymptotic state in equilibrium, one obtains the ideal fluid.
Otherwise, one has to take into account the dissipation term, and this is the
reason that we use the Langevin scheme to derive the evolution equation, but only
on the quantum level. We derive the evolution equation in the integral form
which reveals effects of thermalization.

 Equation (\ref{e20}) is supposed to model all aspects of the deconfinement
or hadronization
processes. The combination $A(\vec{p},t)+ F(\vec{p},t)$ represents the
so-called {\it Langevin force} and is therefore responsible for the
internal dynamics of particle emission in the following manner:
the memory term $A$ causes  dissipation and is
related to stochastic dissipative forces [2]
\begin{equation}
A(\vec{p},t) = \int^{+\infty}_{-\infty}\! d\tau K(\vec{p},t-\tau)
b(\vec{p},\tau), \label{e21}
\end{equation}
with  $K(\vec{p},t)$ being the kernel operator describing the
virtual transitions from one (particle) mode to another.
The operator $F(\vec{p},t)$ is responsible for the action of
a heat bath of absolute temperature $T$ on a particle in the heat bath, and
under the appropriate circumstances is given by
\begin{equation}
F(\vec{p},t) \int^{+\infty}_{-\infty}\!\frac{d\omega}{2\pi}\psi(p_{\mu})\hat{c}(p_{\mu})
e^{-i\omega t} . \label{e22}
\end{equation}
Our heat bath is represented by an ensemble of coupled oscillators,
each described by the operator $\hat{c}(p_{\mu})$ such
that $\left[\hat{c}(p_{\mu}),\hat{c}^{+}(p'_{\mu})\right] \delta^4(p_{\mu}-p'_{\mu})$, and characterized by the noise spectral function
$\psi(p_{\mu})$ [2,1]. Here, the only statistical assumption is that the heat bath
is canonically distributed. The oscillators are coupled to a particle which is
in turn acted upon by an outside force.
Finally, the constant term $P$ in (\ref{e20}) (representing {\it an external source}
term in the Langevin equation) denotes a possible influence of
some external force. This force
would result, for example, in a strong ordering of phases leading,
therefore, to a coherence effect.

The solution of equation (\ref{e20}) is given in $S(\Re_{4})$ by
\begin{equation}
\tilde b(p_{\mu}) = \frac{1}{\omega - \tilde K(p_{\mu})}\,
[\tilde F(p_{\mu}) + \rho (\omega_{P},\epsilon)],
\label{e23}
\end{equation}
where $\omega$ in $\rho (\omega,\epsilon)$ was replaced by the new scale
$\omega_{P} = \omega/P$.
It should be stressed that the term containing
$\rho (\omega_{P},\epsilon)$ as $ \epsilon\rightarrow 0$ gives the general solution to
equation (\ref{e20}). Notice that the distribution $\rho (\omega_{P},\epsilon)$  indicates
the continuous character of the spectrum, and the arbitrary small quantity
$\epsilon$ can be defined by the special physical conditions or the physical
spectra. On the other hand, this $\rho (\omega_{P},\epsilon)$ can be understand as the
temperature-dependent succession (\ref{e4}) where $\epsilon\rightarrow \beta^{-1}$.
Such a succession gives the restriction on the $\beta$-dependent
second term in the solution (\ref{e23}) where at small enough $T$ there will be
a narrow peak at $\omega = 0$.

>From the scattering matrix point of view the solution (\ref{e23}) has the following
physical meaning: at enough outgoing past and future, the fields described by the
operators $\tilde a(p_{\mu})$ are free and, thus, the initial and the final states
of dynamic system are both characterized by the constant amplitudes of states.
Both of these states $\varphi (-\infty)$ and $\varphi (+\infty)$ are related to
each other by some operator $S(\tilde R)$ carrying out the transformation of the state
$\varphi (-\infty)$ to the state $\varphi (+\infty)$ and depending on the behaviour
of $\tilde R(p_{\mu})$:
$$\varphi (+\infty) = \varphi (\tilde R) = S(\tilde R)\varphi (-\infty).$$
In accordance with this definition, it is natural to identify $S(\tilde R)$ as the
scattering matrix in the case of arbitrary sources giving rise to the intensity of
$\tilde R$.

Based on QFT point of view relation (\ref{e11}) indicates the
appearance of the terms containing nonquantum fields which are characterized by
the operators $\tilde R(p_{\mu})$.
Hence, there will be terms with $\tilde R$ in the matrix elements, and
these $\tilde R$ cannot be realized via real particles. The operator function
$\tilde R(p_{\mu})$ could be considered as a limit of an average value of some quantum
operator (or even a set of operators) with intesity increasing up to infinity.
The later statement could be realized in the following mathematical representation [2]:
$$\tilde R(p_{\mu}) = \sqrt {\alpha\,\Xi (p_{\mu},p_{\mu})}, $$
where $\alpha$ is the coherence function which gives the strength of the average
$\Xi (p_{\mu},p_{\mu})$. We shall find this coherence function carrying the
dependence on $\beta$, the particle mass $m$ and the chemical potential $\mu$.

In principal, the interaction with the fields described by $\tilde R$ is provided by
the virtual particles, the process of propagating of which is given by the potentials
defined by the $\tilde R$ operator function.

The condition $M_{ch}\rightarrow 0$ (or $\Omega(r)\sim\frac{1}{M^{4}_{ch}}\rightarrow\infty$)
in the representation
$$\lim_{p_{\mu}\rightarrow p^{\prime}_{\mu}}\Xi (p_{\mu},p^{\prime}_{\mu}) \lim_{Q^{2}\rightarrow 0} \Omega(r)\,n(\bar\omega,\beta)\exp (-q^{2}/2)\rightarrow
\frac{1}{M^{4}_{ch}}\,n(\omega,\beta), $$
with [2]
$$ \Omega(r)=\frac{1}{\pi^2}\,r_{0}\,r_{z}\,r^{2}_{t},\,\,\,
\Xi (p_{\mu},p_{\mu}) = {\langle\tilde a^{+}(p_{\mu}) \tilde a(p_{\mu})\rangle}_{\beta} $$
means that the role of the arbitrary source characterized by the operator function
$\tilde R(p_{\mu})$ in $\tilde b(p_{\mu})= \tilde a(p_{\mu}) + \tilde R(p_{\mu})$ disappears.

Let us go to the thermal field operator $\Phi(x)$ by means of the linear combination
of the frequency parts $\phi^{+}(x)$ and $\phi^{-}(x)$
\begin{equation}
\Phi (x) = \frac{1}{\sqrt{2}}\,\left [\phi^{+}(x) +\phi^{-}(x)\right ]
\label{e24}
\end{equation}
composed of the operators $\tilde b(p_{\mu})$ and $\tilde b^{+}(p_{\mu})$ as the
solutions of equation (\ref{e20}) and conjugate to it, respectively:
\begin{equation}
\phi^{-} (x) = \int \frac{d^{3}p}{(2\pi)^{3} 2 (\vec p^{2} +m^{2})^{1/2}}
\tilde b^{+}(p_{\mu})\,e^{ipx},
\label{e25}
\end{equation}
\begin{equation}
\phi^{+} (x) = \int \frac{d^{3}p}{(2\pi)^{3} 2 (\vec p^{2} +m^{2})^{1/2}}
\tilde b(p_{\mu})\,e^{-ipx}.
\label{e26}
\end{equation}

One can easily find two equations of motion for the Fourier transformed operators
$\tilde b(p_{\mu})$ and $\tilde b^{+}(p_{\mu})$ in $S(\Re_{4})$
\begin{equation}
[\omega - \tilde K(p_{\mu})]\tilde b(p_{\mu}) = \tilde F(p_{\mu}) + \rho(\omega_{P},\epsilon),
\label{e27}
\end{equation}
\begin{equation}
[\omega - \tilde K^{+}(p_{\mu})]\tilde b^{+}(p_{\mu}) \tilde F^{+}(p_{\mu}) + \rho(\omega_{P},\epsilon),
\label{e28}
\end{equation}
which are transformed into new equations for the frequency parts
$\phi^{+} (x)$ and $\phi^{-} (x)$ of the field operator $\Phi (x)$ (\ref{e24})
\begin{equation}
i\partial_{0}\phi^{+} (x) - \int_{\Re_{4}} K(x-y)\,\phi^{+} (y)dy F(x) + P\,\partial_{0}D(x)\,e^{-\epsilon t},
\label{e29}
\end{equation}
\begin{equation}
- i\partial_{0}\phi^{-} (x) - \int_{\Re_{4}} K^{+}(x-y)\,\phi^{-} (y)dy F^{+}(x) + P\,\partial_{0}D(x)\,e^{-\epsilon t}.
\label{e30}
\end{equation}
Here, the field components $\phi^{+}(x)$ and  $\phi^{-}(x)$ are nonlocalized under the effect
of the invariant formfactors $K(x-y)$ and $K^{+}(x-y)$, respectively. In general,
these formfactors can
admit the description of locality for nonlocal interactions. The function $D(x)$  in
(\ref{e29}) and (\ref{e30}) obeys the commutation relation
\begin{equation}
[\Phi (x),\Phi (y)]_{-} = - i D(x)
\label{e31}
\end{equation}
and looks like [9]
\begin{equation}
 D(x)= \frac{1}{2\,\pi}\,\epsilon(x^{0})\,\left [\delta(x^2) -
\frac{m}{2\,\sqrt{x^{2}_{\mu}}}\,\Theta(x^2)\,J_{1}\left (m\sqrt{x^2_{\mu}}\right )\right ],
\label{e32}
\end{equation}
where $\epsilon(x^{0})$ and $\Theta(x^2)$ are the standard unit and the step functions,
respectively, while $ J_{1}(x)$ is the Bessel function. On the mass-shell
$D(x)$ becomes
\begin{equation}
D(x)\simeq \frac{1}{2\,\pi}\,\epsilon(x^{0})\,\left [\delta(x^2) -
\frac{m^2}{4}\,\Theta(x^2) \right ].
\label{e33}
\end{equation}
At this stage, it is needed to stress that we have got new generalized evolution
equations (\ref{e29}) and (\ref{e30}) keeping the general
features of propagating and
interacting of the quantum fields with mass $m$ in the heat bath (reservoir)
and chaotically distorted by  other fields. To proceed to a further analysis,
let us rewrite the system of equations (\ref{e29}) and (\ref{e30}) in the following form:
\begin{equation}
i\partial_{0}\phi^{+} (x) -  K(x)\star\phi^{+} (x) = f(x),
\label{e34}
\end{equation}
\begin{equation}
- i\partial_{0}\phi^{-} (x) -  K^{+}(x)\star\phi^{-} (x) = f^{+}(x),
\label{e35}
\end{equation}
where $A(x)\star B(x)$ means the convoluted function of the generalized functions
$A(x)$ and $B(x)$, and
\begin{equation}
f(x) = F(x) + P\,\partial_{0}D(x)e^{-\epsilon t}.
\label{e36}
\end{equation}
Applying the direct Fourier transformation to both sides of eqs.
(\ref{e34}) and (\ref{e35}) with the following properties of the
Fourier transformation:
\begin{equation}
F[K(x)\star\phi^{+} (x) ] = F[K(x)]F[\phi^{+} (x)],
\label{e37}
\end{equation}
one can get two equations
\begin{equation}
[- p^{0} - \tilde K(p_{\mu})]\tilde\phi^{+}(p_{\mu}) = F[f(x)],
\label{e38}
\end{equation}
\begin{equation}
[p^{0} - \tilde K^{+}(p_{\mu})]\tilde\phi^{-}(p_{\mu}) = F[f^{+}(x)].
\label{e39}
\end{equation}
Multiplying eqs. (\ref{e38}) and (\ref{e39}) by $p^{0} - \tilde K^{+}(p_{\mu})$
and $- p^{0} - \tilde K(p_{\mu})$, respectively, one finds
\begin{equation}
[- p^{0} - \tilde K(p_{\mu})][p^{0} - \tilde K^{+}(p_{\mu})]\tilde\Phi(p_{\mu}) = T(p_{\mu}),
\label{e40}
\end{equation}
where
\begin{equation}
T(p_{\mu}) = [p^{0} - \tilde K^{+}(p_{\mu})]F[f(x)]-
 [p^{0} + \tilde K(p_{\mu})]F[f^{+}(x)].
\label{e41}
\end{equation}
Now we are at the stage of the main strategy:  one should identify the field
$\Phi (x)$ introduced in (\ref{e5}) and the field $\Phi (x)$ (\ref{e24}) built up of the
fields $\phi^{+}$ and $\phi^{-}$ as the solutions of  generalized
equations (\ref{e29}) and (\ref{e30}). The next step is our requirement that the
Green's function
$\tilde G(p_{\mu})$ in (\ref{e10}) and  the function $ \Gamma(p_{\mu},\beta)$,
satisfying eq.(\ref{e40})
\begin{equation}
[p^{0} + \tilde K(p_{\mu})][- p^{0} + \tilde K^{+}(p_{\mu})]\tilde\Gamma(p_{\mu}) = 1,
\label{e42}
\end{equation}
have to be equal to each other, where [9]
\begin{equation}
\tilde G(p_{\mu})\rightarrow \tilde G(p^2, g^2, m^2)\simeq \frac{1 - g^2\,\xi(p^2, m^2)}
{m^2 - p^2 -i\epsilon}
\label{e43}
\end{equation}
with $g$ being the scalar coupling constant and the one-loop correction of the scalar field
$\xi << 1/m^2$ at $1/4 \leq (m^2/p^2)\leq 1$.
It means that we define the operator kernel $\tilde K(p_{\mu})$ in (\ref{e27}) from
the condition of the nonlocal coincidence of the Green's function $\tilde G(p_{\mu})$
in (\ref{e10}) and the thermodynamic function $\tilde\Gamma(p_{\mu},\beta)$
from (\ref{e42}) in $S(\Re_{4})$
$$ \lim_{x-x^{\prime}\sim O(r)} F[\tilde G(p_{\mu}) - \tilde\Gamma(p_{\mu},\beta)] =0.$$

We can easily derive the kernel operator
$\tilde K(p_{\mu})$ in the case of its real nature, i.e.,
\begin{equation}
\tilde K(p_{\mu}) = {(m^2 + \vec p^2)}^{\frac{1}{2}}{\left [ 1 + g^2\xi(p^2, m^2)
\left ( 1- \frac{\omega^2}{m^2 + \vec p^2}\right )\right ]}^{\frac{1}{2}},
\label{e44}
\end{equation}
where
\begin{equation}
\xi(p^2, m^2) = \frac{1}{96\,\pi^2\,m^2}\left (\frac{2\,\pi}{\sqrt{3}} -1\right ),
\,\,\, p^2\simeq m^2,
\label{e45}
\end{equation}
and
\begin{equation}
\xi(p^2, m^2) = \frac{1}{96\,\pi\,m^2}\left (i\, \sqrt {1-\frac{4\,m^2}{p^2}}
+ \frac{\pi}{\sqrt{3}}\right ),
\,\,\, p^2\simeq 4 m^2.
\label{e46}
\end{equation}
The ultraviolet behaviour at $ \vert p^2\vert >> m^2$ leads to

\begin{equation}
\xi(p^2, m^2) \simeq  \frac{-1}{32\,\pi^{2}\,p^2}\left [\ln\frac{\vert p^2\vert}{m^2} -
\frac{\pi}{\sqrt{3}} - i\,\pi\Theta (p^2)\right ].
\label{e47}
\end{equation}

Out of many details (for which we refer to [2,1])  important
in our case is the fact that the $2$-particle BEC function
for like-charge particles is defined as
\begin{eqnarray}
C_2(Q) &=&  \chi(N) \cdot \frac{\tilde{f}(p_{\mu},p'_{\mu})}
{\tilde{f}(p_{\mu})\cdot\tilde{f}(p'_{\mu})}
= \chi(N) \cdot \left[1\, +\, D(p_{\mu},p'_{\mu})\right] ,
\label{e48}
\end{eqnarray}
where $\tilde{f}(p_{\mu},p'_{\mu}) = \langle
\tilde{b}^{+}(p_{\mu})\tilde{b}^{+}(p'_{\mu})
\tilde{b}(p_{\mu})\tilde{b}(p'_{\mu})\rangle$ and
$\tilde{f}(p_{\mu}) = \langle
\tilde{b}^{+}(p_{\mu})\tilde{b}(p_{\mu})\rangle$ are the
corresponding thermal statistical averages with
$\tilde{b}(p_{\mu})$
being the corresponding Fourier transformed solution (\ref{e23}).
The multiplicity $N$ depending factor is equal to
$\chi (N) = \langle N(N-1)\rangle/\langle N\rangle^2$.
As shown in [2], (notice that
operators $\tilde{R}(p_{\mu})$ by definition commute with themselves
and with any other operator considered here):
\begin{eqnarray}
\tilde{f}(p_{\mu},p'_{\mu}) &=& \tilde{f}(p_{\mu})\cdot\tilde{f}(p'_{\mu}) +
\langle\tilde{a}^{+}(p_{\mu})\tilde{a}(p'_{\mu})\rangle
\langle\tilde{a}^{+}(p'_{\mu})\tilde{a}(p_{\mu})\rangle  +
\langle\tilde{a}^{+}(p_{\mu})\tilde{a}(p'_{\mu})\rangle
\tilde{R}^{+}(p'_{\mu})\tilde{R}(p_{\mu})
+\nonumber\\
&+&\langle\tilde{a}^{+}(p'_{\mu})\tilde{a}(p_{\mu})\rangle \tilde{R}^{+}(p_
{\mu})\tilde{R}(p'_{\mu}) ,\label{e49}\\
\tilde{f}(p_{\mu}) &=& \langle \tilde{a}^+(p_{\mu})\tilde{a}(p_{\mu})\rangle
 +\,|\tilde{R}(p_{\mu})|^2 . \label{e50}
\end{eqnarray}
This defines $D(p_{\mu},p'_{\mu}) = \tilde{f}(p_{\mu},p'_{\mu})/
[\tilde{f}(p_{\mu})\cdot\tilde{f}(p'_{\mu})] - 1$ in (\ref{e48}) in
terms of the operators $\tilde{a}(p_{\mu})$ and $\tilde{R}(p_{\mu})$
which in our case are equal to
\begin{equation}
\tilde{a}(p_{\mu}) \frac{\tilde{F}(p_{\mu})}{\omega - \tilde{K}(p_{\mu})}\quad {\rm and}\quad
\tilde{R}(p_{\mu}) \frac{\rho(\omega_{P},\epsilon)}{\omega - \tilde{K}(p_{\mu})} .
\label{e51}
\end{equation}
This means, therefore,
that the correlation function $C_2(Q)$, as defined
by eq. (\ref{e48}), is essentially given in terms of $\rho(\omega_{P},\epsilon)$
and the following two thermal averages for
the thermostat operators $F(\vec{p},t)$:
\begin{eqnarray}
\langle F^{+}(\vec{p},t)F(\vec{p}',t')\rangle \delta^3(\vec{p}-\vec{p}')\,\int \frac{d\omega}{2\pi}\,
               \left|\psi(p_{\mu})\right|^2\, n(\omega)e^{+i\omega(t-t')},
               \label{e52}\\
\langle F(\vec{p},t)F^{+}(\vec{p}',t')\rangle \delta^3(\vec{p}-\vec{p}')\,\int \frac{d\omega}{2\pi}\,
               \left|\psi(p_{\mu})\right|^2\, [1 + n(\omega)] e^{-i\omega(t-t')},
               \label{e53}
\end{eqnarray}
where $n(\omega) = \left\{\exp \left[(\omega - \mu)\beta\right] -
1\right\}^{-1}$ is the number of (by assumption - only bosonic in our
case) oscillators of energy $\omega$ in the reservoir
characterized by parameters $\mu$ (chemical potential) and inverse
temperature $\beta$ [10].
Notice that with only delta functions present
in (\ref{e52}) and (\ref{e53}) one would have a situation in which our
deconfined or hadronizing
system would be described by some kind of {\it white noise} only. The
integrals multiplying these delta functions and depending on $(a)$ momentum
characteristic of a heat bath $\psi(p_{\mu})$ and $(b)$ assumed bosonic statistics
of produced secondaries resulting in factors $n(\omega)$ and $1+n(\omega)$,
respectively, bring the description of the system considered here closer to reality.

It is easy to realize now that the existence of BEC, i.e.,
the fact that $C_2(Q) >1$, is strictly connected with nonzero values
of the thermal averages (\ref{e52}) and (\ref{e53}).
However, in the form
presented there, they differ from zero {\it only at one point},
namely for $Q=0$ (i.e., for $p_{\mu} = p'_{\mu}$). Actually, this is
the price one pays for the QFT assumptions tacitly made here, namely
for the {\it infinite} spatial extension and for the {\it uniformity}
of our reservoir. However, we know from the experiments, e.g., [11,4-6] that
$C_2(Q)$ reaches its maximum at $Q=0$ and falls down towards its
asymptotic value of $C_2 = 1$ at large of $Q$ (actually already at $Q
\sim 1$ GeV/c). To reproduce the same behaviour by means of our
approach, one has to replace the delta functions in eqs.
(\ref{e52}) and (\ref{e53}) by functions with supports
larger than limited to a
one point only. This means that these functions should not be infinite
at $Q_{\mu} = p_{\mu}-p'_{\mu} = 0$ but remain more or less sharply
peaked at this point, otherwise remaining finite and falling to zero
at small but finite values of $|Q_{\mu}|$ (actually the same as
those at which $C_2(Q)$ reaches unity)
\begin{equation}
\delta(p_{\mu} - p'_{\mu})\, \Longrightarrow\, \Omega_0\cdot
\exp [-(p_{\mu}-p^{\prime}_{\mu})L^{\mu\nu}(r)(p_{\nu}-p^{\prime}_{\nu})].
\label{e54}
\end{equation}
Here we  replaced  the $\delta$-function with the smearing (smooth)
dimensionless generalized function
$\Omega(q=Q r)=\exp [-(p_{\mu}-p^{\prime}_{\mu})L^{\mu\nu}(r)(p_{\nu}-p^{\prime}_{\nu})]$,
where the tensor $L^{\mu\nu}$ is defined by the geometry of the space and
$\Omega_0$ has the same dimension as the $\delta$ - function
(actually, it is nothing else but $4$-dimensional volume restricting
the space-time region of particle production).

In this way we tacitly introduce
a new parameter, $r_{\mu}$, a $4$-vector such that
it has dimension of length. This defines the
region of {\it nonvanishing} particle density
with the space-time extension of the
particle emission source. Expression (\ref{e54}) has to be understood
in the sense that $\Omega(Q r)$ is a function
which in the limit of $r\rightarrow \infty$ becomes {\it strictly} a
$\delta$ - function.
With such a replacement one now has
\begin{equation}
D(p_{\mu},p'_{\mu}) \frac{\sqrt{\tilde{\Omega}(q)}}{(1+\alpha)(1+\alpha')}
\cdot \left[ \sqrt{\tilde{\Omega}(q)} + 2\sqrt{\alpha \alpha'}
\right],  \label{e55}
\end{equation}
where
\begin{equation}
\tilde{\Omega}(q) = \gamma \cdot \Omega(q),~~
\gamma = \frac{n^2(\bar{\omega})}{n(\omega)n(\omega')}.
\label{e56}
\end{equation}
The coherence function $\alpha$ is
another very important one which summarizes our knowledge of
other than space-time characteristics of the particle emission source.
Notice that $\alpha > 0$ only when $P \neq 0$. Actually, for $\alpha = 0$ one has
\begin{equation}
1 < C_2(Q) < \chi(N)[1 + \gamma \Omega(Q r)] ,
\label{e57}
\end{equation}
i.e., it is contained between the limits corresponding to very large
(lower limit) and very small (upper limit) values of $P$.
Because of this $\alpha$ plays the role of the {\it coherence}
parameter. Neglecting
the energy-momentum dependence of $\alpha$ and assuming that
$\alpha' = \alpha$ one gets the expression
\begin{equation}
C_2(Q) = \chi(N) \left [1 + \frac{2\alpha}{(1 + \alpha)^2}\cdot
\sqrt{\tilde\Omega(q)}\, +\, \frac{1}{(1+\alpha)^2}\cdot
\tilde\Omega(q)\right ] . \label{e58}
\end{equation}
In fact, since in general $\alpha \neq \alpha'$
(due to the fact that $\omega \neq \omega'$ and, therefore, the number
of states identified here with the number of particles with given
energy $n(\omega)$ is also different) one should rather use the
general form (\ref{e48}) for $C_2$ with details given by
(\ref{e55}) and (\ref{e56}), and with $\alpha$ depending on the hadron mass and
on such characteristics of the emission process as temperature $T$ and
chemical potential $\mu$ occurring in definition of $n(\omega)$.
Notice that eq. (\ref{e58}) differs from the usual
empirical parameterization of $C_2(Q)$ [12,4-6],
\begin{equation}
C_2(Q) \sim [1 + \lambda\cdot \Omega(Q r)][1+ aQ + ...]
, \label{e59}
\end{equation}
which is nothing else but the Goldhaber parameterization [13] at $a=0$
$$C_2(Q) = 1 + \lambda e^{-Q^2\,r^2}$$
with $0< \lambda <1$ being a free parameter adjusting the observed
value of $C_2(Q=0)$ which is customary called "coherence strength factor"
or "chaoticity"  meaning $\lambda =0$ for fully coherent and  $\lambda =1$
for fully incoherent sources; $a$ is a c-number, and with
$\Omega(Q r)$ represented usually as Gaussian.
 Recently eq. (\ref{e59})
has found strong theoretical support expressed in
 detail in [14].
Coming back to the hadron mass dependence of $C_2(Q)$, in particular,
the correlation radius, we find that this dependence comes from
$\alpha$-coherence function (\ref{e1}) containing the operator kernel
$\tilde K(p_{\mu})$ defined correctly up to the second-order of the scalar
coupling constant $g^2$ in (\ref{e44}) in the framework of QFT.

 The $\alpha$ - representation in (\ref{e1}) needs to be clarified.
In fact, $M^{4}_{ch}$ may be  decomposed into two parts: $M^{4}_{ch}\rightarrow
M^{(0)}_{ch}\cdot \bar M^{3}_{ch}$, where $M^{(0)}_{ch}$ is the small massive
scale characterized by the time-like scale $r^0$, while $\bar M_{ch}$ being
the characteristic mass associated with the space inverse components $r_{z},\,r_{t}$,
$\bar M^{3}_{ch}\sim (r_{z} r^{2}_{t})^{-1}$. Taking into account the
properties of the distribution $[\rho (\omega_{P},\epsilon)]^{2}$ one can
suggest the following replacement $M^{4}_{ch}[\rho (\omega_{P},\epsilon)]^{2}
\rightarrow [M^{(0)}_{ch}\rho (0,\epsilon)][\bar M^{3}_{ch}\,\rho (\omega_{P},\epsilon)]$,
where the first multiplier is of the order of $O(1)$ while the second one reflects the
massive scale of the particles with the mass $m$, i.e.,
$\bar M^{3}_{ch}\,\rho (\omega_{P},\epsilon)\sim O(m^2)$. Thus,
\begin{equation}
 \alpha \sim O \left[\frac{m^2}{{\vert\omega -
\tilde{K}(p_{\mu})\vert}^{2}\,n(\omega,\beta)}\right ] .
\label{e60}
\end{equation}

Let us return to the problem of $Q$-dependence of BEC. One more
remark is in order here. The problem with the
$\delta(p_{\mu}-p'_{\mu})$ function encountered in two particle
distributions does not exist in the single particle distributions
which are in our case given by eq. (\ref{e50}) and which can be written as
\begin{equation}
\tilde{f}(p_{\mu})\, = \, (1+\alpha) \cdot \Xi(p_{\mu},p_{\mu}) ,
 \label{e61}
\end{equation}
where $\Xi(p_{\mu},p_{\mu})$ is the one-particle distribution function
for the "free" (undistorted) operator $\tilde{a}(p_{\mu})$, namely
\begin{equation}
\Xi(p_{\mu},p_{\mu})\, =\langle\tilde{a}^+(p_{\mu})\tilde{a}(p_{\mu})\rangle
= \Omega_0 \cdot \left| \frac{\psi(p_{\mu})}
                           {\omega - \tilde{K}(p_{\mu})}\right| ^2
n(\omega) . \label{e62}
\end{equation}
Notice that the actual shape of $\tilde{f}(p_{\mu})$ is dictated by both
 $n(\omega)$ (calculated for fixed temperature
$T$ and chemical potential $\mu$ at energy $\omega$ as given by
the Fourier transform of  field operator $\tilde{K}$  (\ref{e44}) and the
shape of the reservoir in the momentum space provided by
$\psi(p_{\mu})$) and by the $\delta$ - like distribution  of external
force $\rho (\omega_{P},\epsilon)$.
 On the other hand, it is clear from
(\ref{e61}) that $\langle N\rangle = \langle N_{ch}\rangle +
\langle N_{coh}\rangle$ (where $\langle N_{ch}\rangle$ and $\langle
N_{coh}\rangle$ denote multiplicities of particles produced
chaotically and coherently, respectively.
For illustration we plotted in Fig. 1 in detail (with the correct
Gaussian shape of
$\Omega (q)$ function) the dependence of $C_2(Q)$ on different values of
$\alpha = 0,~0.25,~1,~4$
and compared it with
the case when the second term in eq. (\ref{e58}) is neglected,
as is the case of the  majority of phenomenological fits to data.

Finally, we present the lower bound of the particle source size
\begin{equation}
r^2 \geq\frac{1}{Q^2}\ln\frac{\gamma\chi(N)}{(1+\alpha)
(1+\alpha^{\prime})[C_{2}(0)-\chi(N)]}
 \label{e63}
\end{equation}
carrying the dependence of: \\
- the absolute temperature of a heat bath
and the chemical potential ($\gamma$-factor defined in (\ref{e56})),\\
- the mean multiplicity  factor $\chi(N)$ defined within formula (\ref{e48})),\\
- the maximal value of the BEC function $C_{2}(Q=0)$ at the origin under
the following condition:
\begin{equation}
C_{2}(0)\leq\chi(N)\left [1+\frac{\gamma}{(1+\alpha)
(1+\alpha^{\prime})}\right ],
 \label{e64}
\end{equation}
- and the hadron mass $m$ given by $\alpha$ and $\alpha^{\prime}$ defined in approximation
(\ref{e60}) with the kernel operator $\tilde K(P_{\mu})$ (\ref{e44}) calculated in QFT.

To summarize: using the QFT supplemented by Langevin-like evolution equations
(\ref{e29}) and  (\ref{e30}) to describe deconfinement or hadronization
processes we have derived the two-particle BEC function in the form explicitly showing the origin of
both the so-called coherence (and how it influences the structure of
BEC) and the $Q$-dependence of BEC represented by the correlation
function $C_2(Q)$. The dynamic source of coherence is identified in
our case with the existence of a constant external term $P$ in the
evolution equation.
 Therefore, for $P\rightarrow \infty$ we have all phases
aligned in the same way and $C_2(Q) =1$. This is because the coherence
has already been introduced on the
level of a particle production source as a property of fields
or operators describing produced particles.
It is therefore up to the experiment to decide which
proposition is followed by nature: the simpler formula
(\ref{e59}) or rather the more involved (\ref{e48}) together
with (\ref{e55}).
>From Fig. 1
one can see that
differences between both forms are clearly visible, especially for
larger values of the coherence function $\alpha$.
\begin{figure}
\noindent
\centerline{\epsfig{file=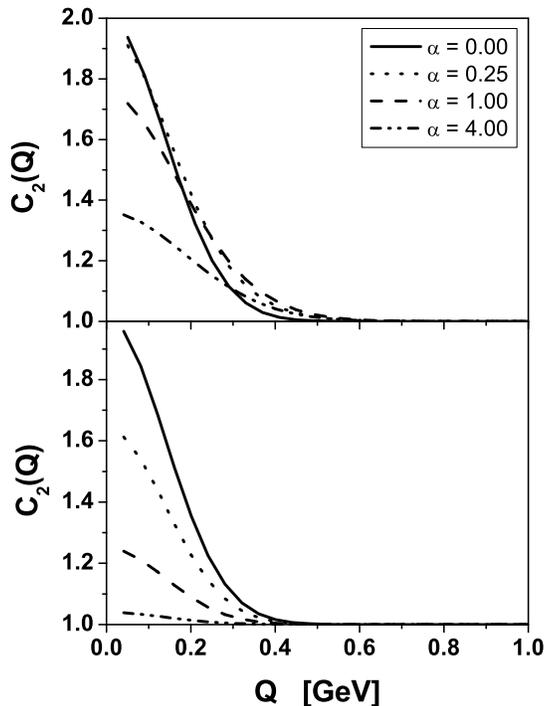, width= 72mm}}
\caption{Shapes of $C_2(Q)$ as given by eq. (\protect\ref{e58}) -
upper panel and for the truncated version of (\protect\ref{e58})
(without the middle term) - lower panel. Gaussian shape of $\Omega(q)$ was
used in both the cases.}
\label{fig:plot2}
\end{figure}
>From our approach it is also clear that the form of $C_2$
reflects distributions of the space-time separation between the
two observed particles.

Finally, we would like to stress that our discussion is so far
limited to only a single type of secondaries being produced. It is
also aimed at a description of deconfinement or hadronization understood as kinetic
freeze-out in some more detailed approaches.
This is enough to obtain our general
goals, i.e., to explain the possible dynamic origin of coherence in
BEC, the origin of the specific shape of the correlation $C_2(Q)$
functions and to explain the dependence of the correlation radius on the
hadron mass which is carryed out by the coherence function $\alpha$,
as seen from the QFT perspective.
It is then plausible that in the general description of the BEC effect
they should be somehow combined, especially if experimental data
indicate such a necessity.

Part of this work is based on a collaboration with G. Wilk and O. Utuyzh.
I have greatly benefited from our many animated discussions. I am also grateful
to S. Tokar for his kind hospitality and for many fruitful discussions.


\begin{references}
\bibitem{10}

The origin of the two parameters occurring at this point,
                 the temperature $\beta =1/T$ and chemical potential $\mu$
                 is the Kubo-Martin-Schwinger condition that
                 $\langle a(\vec{p}',t')a^{+}(\vec{p},t)\rangle                  \langle a^{+}(\vec{p},t)a({p}',t-i\beta)\rangle
                 \cdot \exp(-\beta \mu)$, see [2].
\end{references}
\end{document}